\documentclass[prl,twocolumn]{revtex4}

\usepackage{amssymb,amsmath,amsthm,graphics,graphicx}
\usepackage{hyperref} 
\usepackage{latexsym}
\usepackage{bm}
\allowdisplaybreaks

\def\be{\begin{equation}}
\def\ee{\end{equation}}
\def\beq{\begin{eqnarray}}
\def\eeq{\end{eqnarray}}

\newcommand{\eg}{{\it e.g.,}\ }
\newcommand{\ie}{{\it i.e.\,}\ }

\graphicspath{{Figs/}} 

\begin{document}

\def\lsim{\mathrel{\rlap{\lower4pt\hbox{\hskip1pt$\sim$}}
    \raise1pt\hbox{$<$}}}
\def\gsim{\mathrel{\rlap{\lower4pt\hbox{\hskip1pt$\sim$}}
    \raise1pt\hbox{$>$}}}
\def\be{\begin{equation}}
\def\ee{\end{equation}}
\def\bea{\begin{eqnarray}}
\def\eea{\end{eqnarray}}
\newcommand{\dd}{\mathrm{d}}
\newcommand{\LL}{\mathcal{L}}
\newcommand{\DD}{\mathcal{D}}

\title{Eigenvalue repulsions in the quasinormal spectra of the Kerr-Newman black hole}


\author{\'Oscar J.~C.~Dias}
\affiliation{STAG research centre and Mathematical Sciences, University of Southampton, UK}

\author{Mahdi Godazgar}
\affiliation{School of Mathematical Sciences, Queen Mary University of London, Mile End Road, London E1 4NS, UK.}

\author{Jorge E.~Santos}
\affiliation{DAMTP, Centre for Mathematical Sciences,
    University of Cambridge, Wilberforce Road, Cambridge CB3 0WA, United Kingdom}

\author{Gregorio Carullo}
\affiliation{Dipartimento di Fisica ``Enrico Fermi'', Universit\`a di Pisa, Pisa I-56127, Italy}
\affiliation{INFN sezione di Pisa, Pisa I-56127, Italy}

\author{Danny Laghi}
\affiliation{Dipartimento di Fisica ``Enrico Fermi'', Universit\`a di Pisa, Pisa I-56127, Italy}
\affiliation{INFN sezione di Pisa, Pisa I-56127, Italy}
\affiliation{Laboratoire des 2 Infinis - Toulouse (L2IT-IN2P3), Universit\'e de Toulouse, CNRS, UPS, F-31062 Toulouse Cedex 9, France}

\author{Walter Del Pozzo}
\affiliation{Dipartimento di Fisica ``Enrico Fermi'', Universit\`a di Pisa, Pisa I-56127, Italy}
\affiliation{INFN sezione di Pisa, Pisa I-56127, Italy}

\begin{abstract}
We study the gravito-electromagnetic perturbations of the Kerr-Newman (KN) black hole metric and identify the two $-$  photon sphere and near-horizon $-$ families of quasinormal modes (QNMs) of the KN black hole, computing the frequency spectra (for all the KN parameter space) of the modes with the slowest decay rate. We uncover a novel phenomenon for QNMs that is unique to the KN system, namely eigenvalue repulsion between QNM families. Such a feature is common in solid state physics where \eg it is responsible for energy bands/gaps in the spectra of electrons moving in certain Schr\"odinger potentials. 
Exploiting the enhanced symmetries of the near-horizon limit of the near-extremal KN geometry we also develop a matching asymptotic expansion that allows us to solve the perturbation problem using separation of variables and provides an excellent approximation to the KN QNM spectra near extremality.
The  KN QNM spectra here derived are required not only to account for the gravitational emission in astrophysical environments, such as the ones probed by LIGO, Virgo and LISA, but also allow to extract observational implications on several new physics scenarios, such as mini-charged dark-matter or certain modified theories of gravity, degenerate with the KN solution at the scales of binary mergers.
\end{abstract}

\maketitle


\emph{\bf  Introduction.}

The black hole (BH) uniqueness theorems single out the Kerr-Newman (KN) metric as the most 
general regular, stationary and asymptotically flat electro-vacuum solution of Einstein-Maxwell’s equations~\cite{Mazur}.
Nevertheless, astrophysical BHs are not expected to be able to retain a significant amount of electric charge~\cite{Gibbons, Znajek}.
Consequently, all LIGO-Virgo~\cite{LIGO,Virgo} observations of events compatible with BH binaries~\cite{O3a_catalog} have been so far described 
under the assumption that the merging objects can be modelled by the Kerr metric, the zero-charge limit of the KN solution.   
Due to the lack of template models describing coalescing KN BHs (especially in the merger-ringdown regime),
the zero-charge assumption has not yet been verified in full on observational data,
although see Refs.~\cite{Gupta:2021rod, Bozzola:2020mjx} for recent work in this direction.
Gravitational-waves (GWs) observations of BH mergers are now probing the largest curvature regimes ever reached, 
enabling the experimental study of gravity in its strong-field and dynamical regime~\cite{O3a_catalog, O3a_TGR}
and opening an observational window on potential unobserved gravitational phenomena.
Here, we further the characterisation of KN solutions by finding the full gravito-electromagnetic 
quasinormal mode (QNM) spectrum of KN BHs. The determination of the QNM spectrum
requires solving a coupled system of two partial differential equations (PDEs) for two gauge invariant Newman-Penrose (NP) fields~\cite{Dias:2015wqa} that, 
upon gauge fixing, reduce to the PDE system originally found by Chandrasekhar~\cite{Chandra:1983,Mark:2014aja}.  
Since the publication of Chandrasekhar's seminal work~\cite{Chandra:1983}, despite several attempts,
this task has remained a major open problem in Einstein-Maxwell theory for the past 40 years.

Perturbative results in the small rotation parameter $a$~\cite{Pani:2013ija,Pani:2013wsa} and in small charge parameter 
$Q$~\cite{Mark:2014aja} expansions about the Reissner-Nordstr\"om (RN) and Kerr backgrounds are available.
Ref.~\cite{Dias:2015wqa} did a numerical search of KN modes that could eventually develop an instability but found none, 
thus providing evidence for the linear mode stability of KN (further supported by the non-linear time evolution study of~\cite{Zilhao:2014wqa}). 
In this Letter, motivated also by applications in both ground and space-based GW 
detectors~\cite{LIGO,Virgo,Kagra,ET,CE,LISA}, we instead identify all the gravito-electromagnetic QNM families of the KN 
BH and compute the frequency spectra (across the full KN parameter space) of the most dominant modes, \ie the ones with slowest decay. 
These are the modes that reduce $-$in Chandrasekhar's notation~\cite{Chandra:1983}$-$ to the $Z_2$, $\ell=m=2$, $n=0$ modes 
in the Schwarzschild limit ($a=Q=0$), where the harmonic number $\ell$ gives the number of zeros of the eigenfunction along 
the polar direction and $n$ is the radial overtone. Remarkably, we find that the KN frequency spectra $-$ unlike its $a=0$ 
and/or $Q=0$ limits $-$ are populated with intricate phenomena known as {\it eigenvalue repulsions}. 
The observational applications of our results are not limited to modelling the GW emission in realistic astrophysical environments,
but include the possibility of constraining certain dark matter~\cite{2016JCAP...05..054C} and modified gravity~\cite{Bozzola:2020mjx} models.
The full implications of these results to GW observations are explored in a companion paper~\cite{AstroConstraints}.

\emph{\bf  Formulation of the problem.}
The KN BH solution can be described in standard Boyer-Lindquist coordinates $\{t,r,\theta,\phi\}$ (time, radial, polar, azimuthal coordinates)~\cite{Adamo:2014baa}. The Killing vector $K=\partial_t +\Omega_H \partial_\phi$ generates the event horizon with angular velocity $\Omega_H$ and temperature $T_H$. The event horizon location $r_+$ is the largest root of the function $\Delta$. In terms of the mass, rotation, and charge parameters $\{M,a,Q\}$, these quantities are:  
\begin{eqnarray}\label{KNsol}
&&\hspace{-0.4cm} \Delta = r^2 -2Mr+a^2+Q^2,\ \  r_\pm=M\pm\sqrt{M^2-a^2-Q^2},  \nonumber \\
&&\hspace{-0.4cm}  \Omega_H= \frac{a}{r_+^2+a^2} \,, \qquad 
T_H = \frac{1}{4 \pi  r_+}\frac{r_+^2-a^2-Q^2}{r_+^2+a^2 }. 
\end{eqnarray}
At $r_-=r_+$, \ie  $a=a_{\hbox{\footnotesize ext}}=\sqrt{M^2-Q^2}$, the KN BH has a regular extremal (``ext") configuration with $T_H^{\hbox{\footnotesize ext}} =0$, and maximum angular velocity $\Omega_H^{\hbox{\footnotesize ext}} =a_{\hbox{\footnotesize ext}}/(M^2+a_{\hbox{\footnotesize ext}}^2)$.

Since $\partial_t,\partial_\phi$ are Killing vector fields of KN, its gravito-electromagnetic perturbations can be Fourier decomposed as $e^{-i \omega t} e^{i m \phi}$, where  $\omega$ and $m$ are the frequency and azimuthal quantum number of the mode. Using the NP formalism,~\cite{Dias:2015wqa} derived a set of two coupled PDEs for two gauge invariant quantities $\psi_{-2}$ and $\psi_{-1}$ that describe the most general perturbations (except for trivial modes that shift the parameters of the solution) of a KN BH, namely:
\begin{eqnarray}\label{ChandraEqs}
&& \left(\mathcal{F}_{-2}+ Q^2 \mathcal{G}_{-2}\right) \psi_{-2}  + Q^2 \mathcal{H}_{-2}  \psi_{-1} =0 \,,  \nonumber\\
&& \left(\mathcal{F}_{-1} +Q^2 \mathcal{G}_{-1}\right)\psi_{-1} + Q^2  \mathcal{H}_{-1} \psi_{-2}=0  \,, 
\end{eqnarray}
where the second order differential operators $\{\mathcal{F},\mathcal{G},\mathcal{H}\}$ are in Eq.\ \eqref{def:opsFGH} of the Supplemental Material. The gauge invariant (under diffeomorphisms and NP tetrad rotations) perturbed quantities $\psi_{-2}$ and $\psi_{-1}$ are a combination of NP scalars  $\Psi$'s and $\Phi$'s (see the Supplemental Material).

To solve the coupled PDEs \eqref{ChandraEqs}, we need to impose physical boundary conditions. 
At spatial infinity, we require only outgoing waves, and at the future event horizon, we keep only regular modes in ingoing Eddington-Finkelstein coordinates. Finally, we must require regularity at the North (South) pole $\theta =\pi\,(-\pi)$. See the Supplemental Material for more details.

A scaling symmetry of the system allows us to work with the adimensional parameters $\{\tilde{a},\tilde{Q},\tilde{\omega}\}\equiv \{a/M,Q/M,\omega M\}$ (or $\{\hat{a},\hat{Q},\hat{\omega}\}\equiv\{a/r_+,Q/r_+,\omega r_+\}$). The $t -\phi$ symmetry of KN means that we need only consider modes with  Re$(\omega)\geq 0$, as long as we study both signs of $m$ \footnote{When $a=0$ this enhances to a $t \to -t $ symmetry and the QNM frequencies form pairs of $\{\omega,-\omega^* \}$.}. To solve the PDE problem numerically, we use a pseudospectral method that searches directly for specific QNMs using a Newton-Raphson root-finding algorithm. We refer to the review~\cite{Dias:2015nua} and ~\cite{Dias:2009iu,Dias:2010eu,Dias:2010maa,Dias:2010gk,Dias:2011jg,Dias:2010ma,Dias:2011tj,Cardoso:2013pza,Dias:2014eua,Dias:2018etb} for details. The exponential convergence of the method, and the use of quadruple precision, guarantee that the results are accurate up to, at least, the eighth decimal place.

\begin{figure*}
\includegraphics[width=.32\textwidth]{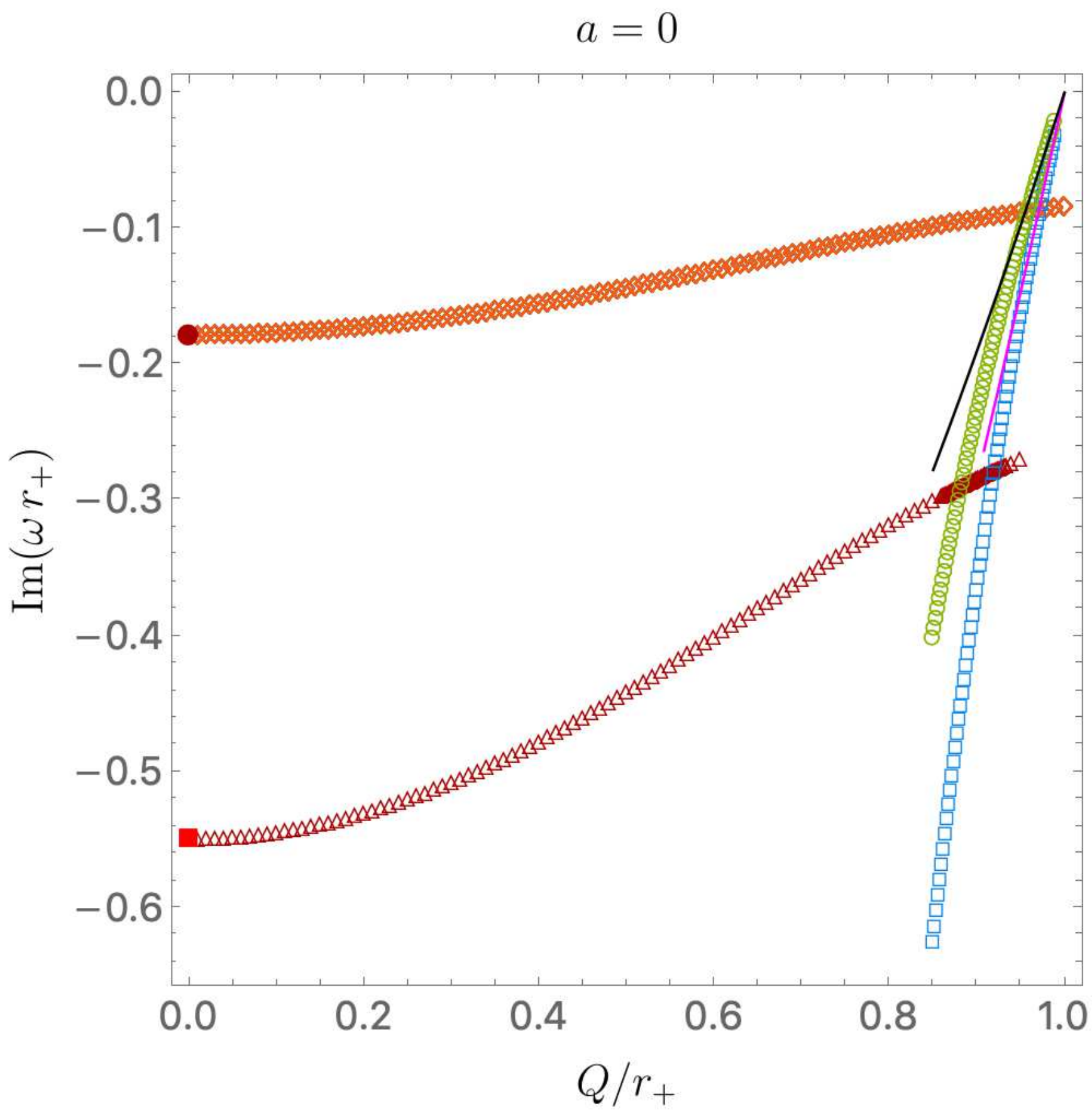}
\includegraphics[width=.32\textwidth]{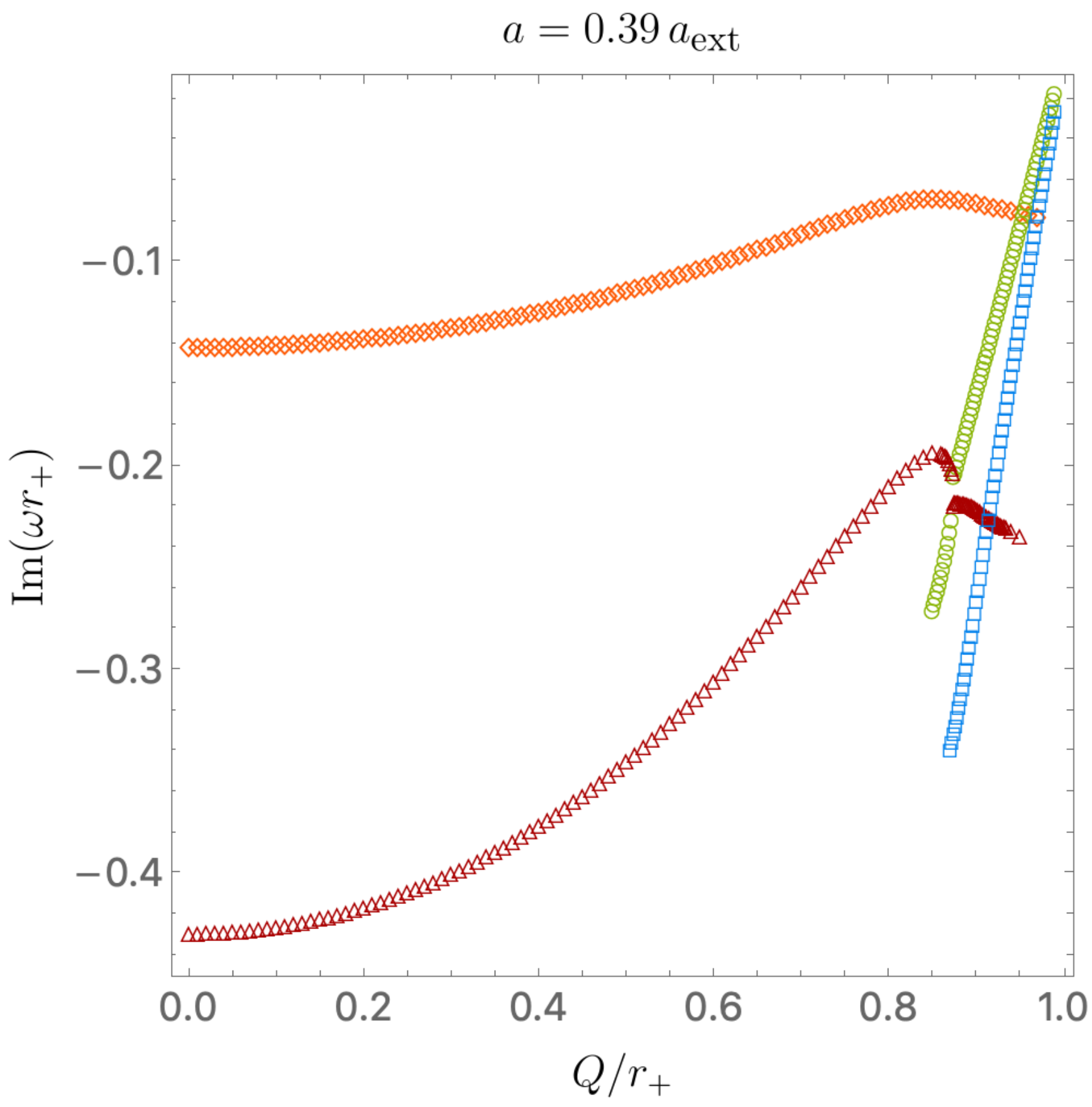}
\includegraphics[width=.33\textwidth]{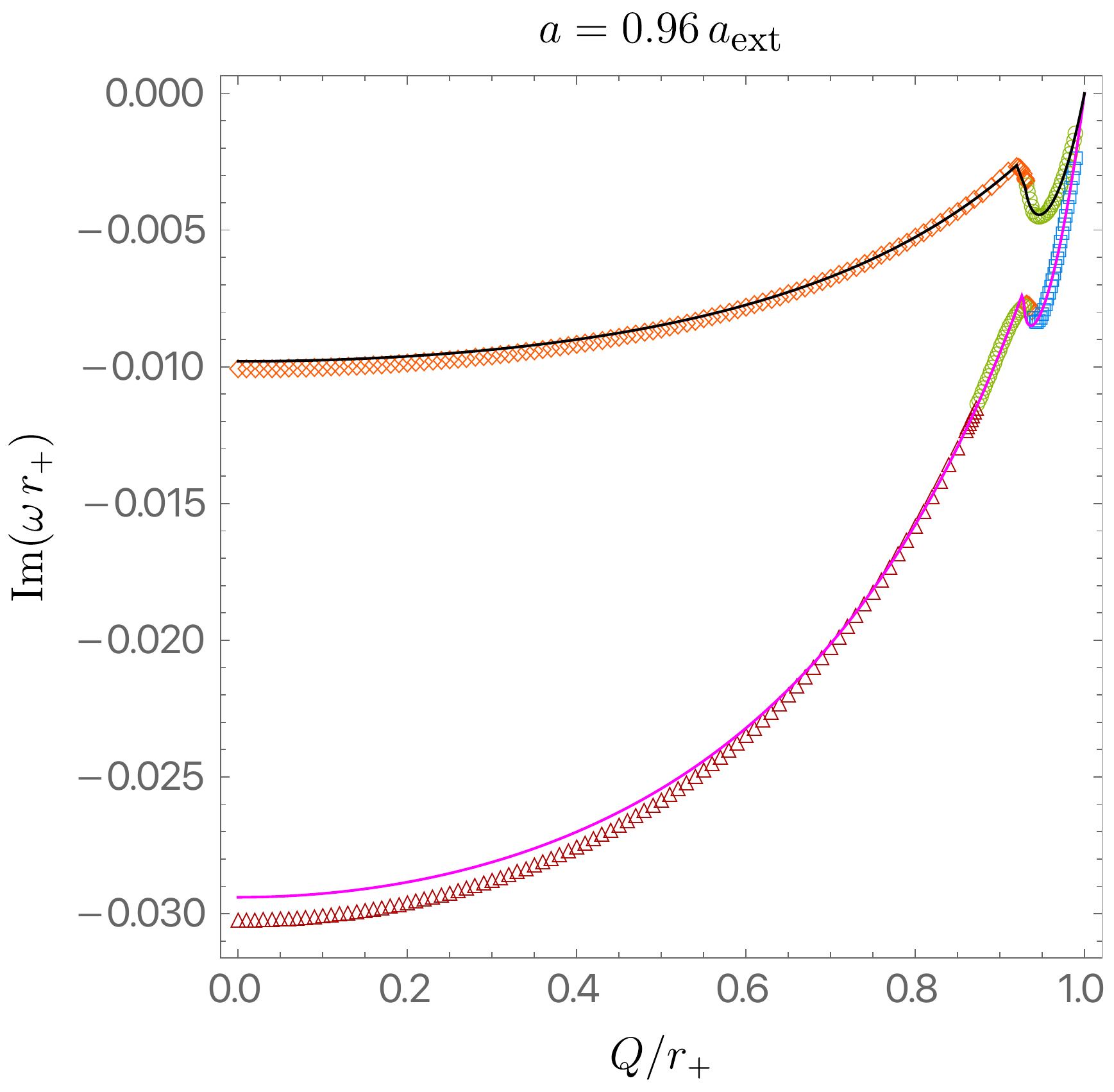}
\caption{QNM spectra for KN BHs with $a/a_{\hbox{\footnotesize ext}}=0$ (left), $0.39$ (middle) and $0.96$ (right). In the RN case, there is an unambiguous QNM family classification: the orange diamond (dark-red triangle) curve is the $n=0$ ($n=1$) PS family which reduces to the dark-red disk $\omega\, r_+=0.74734337 - 0.17792463\, i $ (red square $\omega \, r_+=0.69342199 - 0.54782975\, i$) in the Schwarzschild limit ~\cite{Chandra:1983,Leaver:1985ax}. The green circle (blue square) curve is the $n=0$ ($n=1$) NH family (not shown: for $\hat{Q}<0.85$ these curves extend to lower $\mathrm{Im}\,\hat{\omega}$). In the middle and right panels one observes eigenvalue repulsions unique to the KN QNM spectra. On the left/right panels we also show the frequency  $\tilde{\omega}_{\hbox{\tiny NH}}^{\hbox{\tiny MAE}}$ given by \eqref{NH:freq} for $n=0$ (black solid curve) and for $n=1$ (magenta solid curve).
}
\label{Fig:spectraFix-a}
\end{figure*}

\emph{\bf Analytical analysis and eigenvalue repulsion.}
There are regimes of the parameter space where the frequency of the QNMs can be well approximated by analytical formulae obtained from perturbation/WKB expansions.
This helps identify different families of QNMs. There are two main families of QNMs: 1) the {\it photon sphere} (PS), and 2) the {\it near-horizon} (NH) families. However, as we will find later,  this sharp distinction is unambiguous only for small values of the rotation parameter. In particular, we can see this clearly for the $a= 0$ Reissner-Nordstr\"om (RN) case, the imaginary part of the frequency spectra of which is shown in the left panel of Fig.~\ref{Fig:spectraFix-a} (in units of $r_+$ since some curves change too much in a small range of charge if we use units of $M$) \footnote{The frequency spectra in the left panel of Fig.~\ref{Fig:spectraFix-a} was obtained solving the coupled pair of KN PDEs and, independently, the Regge-Wheeler$-$Zerilli ODE~\cite{Regge:1957td,Zerilli:1974ai} that describes perturbations of RN. The fact that both match validates our numerics for the KN PDEs. See \cite{Berti_2009} for a detailed review on QNM studies of the RN and Kerr black holes.}. Letting $n=0,1,\cdots$ denote the radial overtone, the orange diamond and dark-red triangle curves describe the $n=0$ ($\mathrm{PS}_{0}$) and $n=1$ ($\mathrm{PS}_{1}$) PS families, respectively. And, the green circle and  blue square curves describe the $n=0$ ($\mathrm{NH}_{0}$) and $n=1$ ($\mathrm{NH}_{1}$) NH families. Focusing our attention on the families with slowest decay rate, the $\mathrm{PS}_{0}$ and $\mathrm{NH}_{0}$ curves intersect (simple crossover) at  
$\hat{Q}= \hat{Q}_c^{\hbox{\tiny RN}} \simeq 0.959227$ ($\tilde{Q}\equiv \tilde{Q}_c^{\hbox{\tiny RN}} \simeq 0.9991342$). For  $0 \leq \hat{Q}<\hat{Q}_c^{\hbox{\tiny RN}}$, $\mathrm{PS}_{0}$ is the dominant QNM, while for $\hat{Q}_c^{\hbox{\tiny RN}} \leq \hat{Q}\leq 1$ it is the $\mathrm{NH}_{0}$ QNM that has smaller $|\mathrm{Im}\,\hat{\omega}|$.

In the eikonal or geometric optics limit (the WKB limit $\ell\sim |m| \gg 1$) the PS QNM  frequencies  are known to be related to the properties of the equatorial plane unstable circular photon orbits. 
The real and imaginary parts of the PS frequency are proportional to the Keplerian frequency $\Omega_c$  and to the Lyapunov exponent $\lambda_L$, respectively~\cite{Goebel:1972,Ferrari:1984zz,Ferrari:1984ozr,Mashhoon:1985cya,Bombelli:1991eg,Cornish:2003ig,Cardoso:2008bp,Dolan:2010wr,Yang:2012he,Stuchlik1991}.
The latter describes how quickly a null geodesic congruence around the orbit increases its cross section under radial deformations. In this limit, the PS frequencies are (see~\cite{Zimmerman:2015trm} and  \cite{ExtendedQNMsKN})
\be\label{PS:eikonal}
 \omega^{\hbox{\tiny eikn}}_{\hbox{\tiny PS}}  \simeq  \frac{m}{b_s} -i\,\frac{n+1/2}{b_s r_s^2}\,\frac{\left|r_s^2+a^2-a b_s\right|}{|b_s-a|\left(6 r_s^2+a^2-b_s^2\right) ^{-\frac{1}{2}}}\,, 
\ee
where $r_s$ and $b_s$ are the radius and impact parameter of the unstable orbits defined implicitly in terms of  $M$, $Q$:
\be\label{PS:MQrsbs}
M=\frac{r_s \left(b_s^2-a^2-2 r_s^2\right)}{\left(b_s-a\right)^2}\,, \quad Q=\frac{r_s\sqrt{b_s^2-a^2-3 r_s^2}}{\sqrt{\left(b_s-a\right){}^2}}.
\ee
There are two real roots $r_s$ higher than $r_+$ which are in correspondence with two PS modes: the {\it co-rotating} one (with $m=\ell$) that maps to the eikonal orbit with radius $r_s=r_s^-$ and $b_s>0$ (and that has the lowest $|\mathrm{Im}\,\tilde{\omega}|$) and the {\it counter-rotating} mode with $m=-\ell$ which is in correspondence with the orbit with radius $r_s=r_s^+$ and $b_s<0$, with $r_s^+\geq r_s^- \geq r_+$. As a check, we find that \eqref{PS:eikonal} is in excellent agreement with the numerical data for $\ell=m=6$ (see  \cite{ExtendedQNMsKN}), and it still gives a reasonable approximation when $\ell=m=2$. Altogether, this identifies the PS QNM family and validates our numerics.  

Now let us discuss the NH family of QNMs. In the RN case (left panel of Fig.~\ref{Fig:spectraFix-a}),  this is the dominant QNM near extremality, \ie for $\hat{Q}_c^{\hbox{\tiny RN}} <\hat{Q}\leq 1$. 
Near extremality, the RN and KN wavefunctions are very localized near the horizon and quickly decay to zero away from it. This suggests doing a  poor-man's matching asymptotic expansion (MAE), whereby we take the {\it near-horizon limit} of the perturbed equations \eqref{ChandraEqs} to find the near-region solution (which we solve analytically) and match with a {\it vanishing} far-region wavefunction in the overlapping region where both solutions are valid  \footnote{Ideally, we would also solve the far-region equations to obtain the sub-leading far-region solution but in the KN background we cannot do it analytically.}. In fact, motivated by the result that the near-horizon limit of the extremal KN BH corresponds to a warped circle fibred over $AdS_2$ (Anti-de Sitter)~\cite{Bardeen:1999px}, the perturbations of which can be decomposed as a sum of known radial  $AdS_2$ harmonics, we can use {\it separation of variables}.  Therefore, the system of 2 coupled PDEs for $\{\psi_{-2},\psi_{-1}\}$ separates into a system of 2 decoupled radial ordinary differential equations (ODEs) and a coupled system of 2 angular ODEs. This yields an analytical expression for the NH frequency (derivation is given in \cite{ExtendedQNMsKN}):
\begin{equation}\label{NH:freq}
 \tilde{\omega}_{\hbox{\tiny NH}}^{\hbox{\tiny MAE}} \simeq  
 \frac{m\tilde{a}}{1+\tilde{a}^2}
+\sigma \bigg[\frac{m \tilde{a}(1-\tilde{a}^2)}{2(1+ \tilde{a}^2)^2}-\frac{i}{4} \frac{1+2n}{1+ \tilde{a}^2}-\,\frac{\sqrt{-\lambda _2}}{4(1+\tilde{a}^2)^2}\bigg] 
\end{equation}
where $n=0,1,2,\cdots$ is again the radial overtone,  $\tilde{a}=\tilde{a}_{\hbox{\footnotesize ext}}$, and the expansion is over the off-extremality parameter $\sigma=1-\frac{r_-}{r_+}$ up to $\mathcal{O}\left(\sigma^2\right)$. Here, $\lambda_2(m,\tilde{a}_{\hbox{\footnotesize ext}})$ is a separation constant that we find by solving numerically the aforementioned coupled system of two angular ODEs. In our conventions $\mathrm{Re}(\sqrt{z})>0$ and $\mathrm{Im}(\sqrt{z})>0$ when $z$ is positive and negative, respectively. Our initial derivation of \eqref{NH:freq} is valid for $\lambda_2>0$ but, motivated by the Kerr results reported in~\cite{Yang:2012pj,Yang:2013uba},  we will use it also when $\lambda_2<0$. In a complementary manner, in the WKB limit $m\gg 1$, $\lambda_2$ is well approximated by
\be \label{NH:lambda2wkb}
\lambda_2^{\hbox{\tiny WKB}}=\lambda_{2,0}\,m^2+\lambda_{2,1}\,m+\lambda_{2,2}+\frac{\lambda_{2,3}}{m}+\mathcal{O}\left(1/m^2\right),
\ee
where the WKB coefficients $\lambda_{2,0},\cdots,\lambda_{2,4}$ are functions of $\tilde{a}$ given in Eq.\ \eqref{NH:WKBansatzCoef} of the Supplemental Material. At extremality ($\sigma=0$), \eqref{NH:freq} reduces to $\mathrm{Re}\, \tilde{\omega}=m \tilde{\Omega}_H^{\hbox{\footnotesize ext}}$ and $\mathrm{Im} \,\tilde{\omega}=0$, and in the Kerr and RN limits, it reduces to the expressions first found in ~\cite{Yang:2012pj,Yang:2013uba} and~\cite{Zimmerman:2015trm}, respectively.
\begin{figure*}
\includegraphics[width=.4\textwidth]{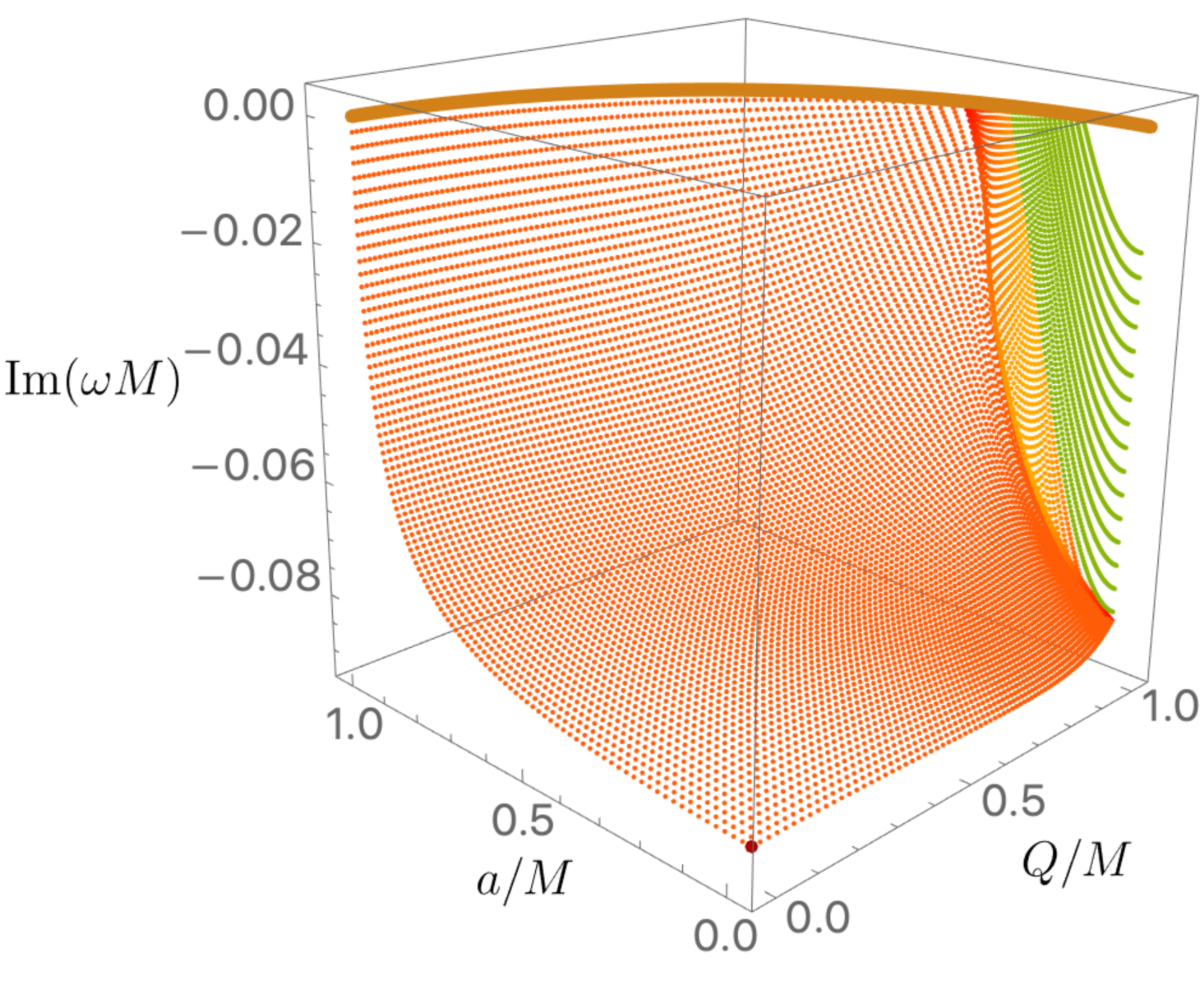}
\hspace{1.5cm}
\includegraphics[width=.38\textwidth]{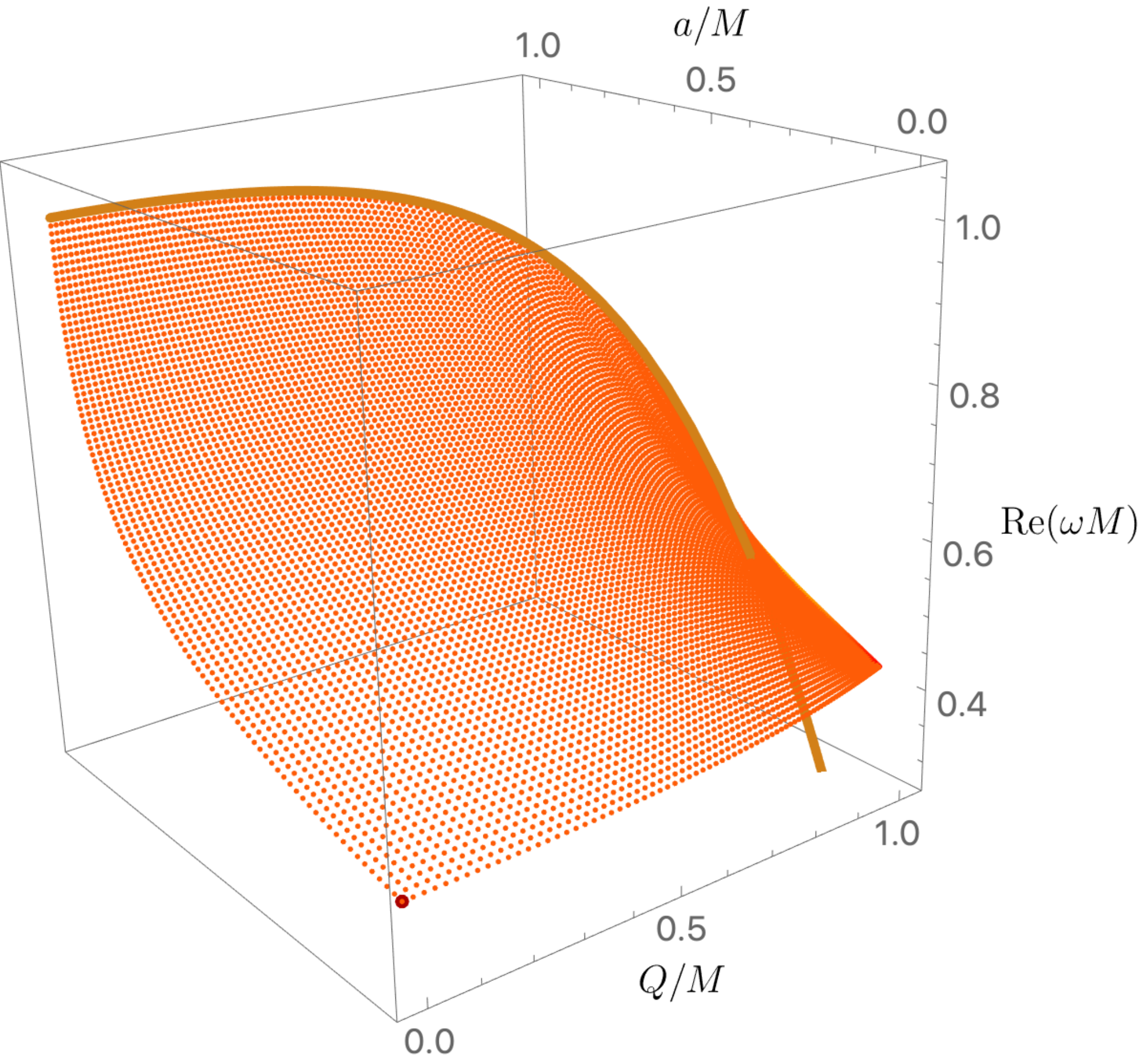}
\caption{Imaginary (left panel) and real (right panel) parts of the frequency for the $Z_2$, $\ell=m=2, n=0$ KN QNM with lowest $\mathrm{Im}\,|\tilde{\omega}|$. At extremality, the dominant mode always starts at $\mathrm{Im}\,\tilde{\omega}=0$ and $\mathrm{Re}\,\tilde{\omega}=m\tilde{\Omega}_H^{\hbox{\footnotesize ext}}$ (brown curve). The dark-red point ($a=0=Q$), $\tilde{\omega}\simeq 0.37367168 - 0.08896232\, i $, is the gravitational QNM of Schwarzschild ~\cite{Chandra:1983,Leaver:1985ax}. In the right panel, the orange and green regions are so close to the extremal brown curve that they are not visible.}
\label{Fig:Z2l2m2n0}
\end{figure*}  

Approximation \eqref{NH:freq} is in excellent agreement with the numerical frequencies (near extremality). This is illustrated in the left and right panels of Fig.~\ref{Fig:spectraFix-a}. For the RN case (left panel), extremality is at $\hat{Q}=1$ and \eqref{NH:freq} with $n=0$ (black line) gives the correct slope for the $\mathrm{NH}_0$ family (green circles), while \eqref{NH:freq} with $n=1$ (magenta line) yields the slope of the $\mathrm{NH}_1$ family (blue squares). On the right panel, we take a KN BH family with $a/a_{\hbox{\footnotesize ext}}=0.96$ (so the whole family of solutions is close to extremality) and compare the numerical results for the dominant $n=0$ QNMs (curve that connects orange diamonds and green circles) with the black  curve, \ie \eqref{NH:freq} with $n=0$. 
Moreover, we also compare  \eqref{NH:freq} with $n=1$ (magenta curve) with the $n=1$ numerical modes with the second slowest decay rate (3-branched curve connecting the dark-red triangles, green circles and  blue squares).
So, \eqref{NH:freq} clearly identifies the NH family in the RN limit, and more generically, the dominant modes near extremality. 

The right panel of  Fig.~\ref{Fig:spectraFix-a} illustrates a remarkable property of KN QNMs. In the RN case and for small rotation, the  $\mathrm{PS}_0$ family dominates the spectra for  $0 \leq \tilde{Q}<\tilde{Q}_c(\tilde{a})$ (with $\tilde{Q}_c(0)=\tilde{Q}_c^{\hbox{\tiny RN}}$) while the  $\mathrm{NH}_0$ family dominates for $\tilde{Q}_c(\tilde{a})<\tilde{Q}\leq 1$.
But, when  $\tilde{a}$ grows and approaches to extremality, at $a/a_{\hbox{\footnotesize ext}}=0.96$, the $\mathrm{PS}_0$ family {\it merges} with the $\mathrm{NH}_0$ family (orange diamond and green circle curves merge in the right panel of Fig.~\ref{Fig:spectraFix-a}). For higher $a/a_{\hbox{\footnotesize ext}}$ the two families remain merged and this line of solutions approaches $\mathrm{Im} \,\tilde{\omega}=0$, $\mathrm{Re}\, \tilde{\omega}=m \tilde{\Omega}_H^{\hbox{\footnotesize ext}}$ as $a\to a_{\hbox{\footnotesize ext}}$. The whole $n=0$ QNM curve in the right plot is thus well approximated by \eqref{NH:freq}: it captures the $\mathrm{NH}_0$ modes in the RN limit but also the ``$\mathrm{PS}_0$-$\mathrm{NH}_0$ merged" modes (when close to extremality). 

The above features of the KN QNMs can be best understood in terms of a critical rotation $\tilde{a}_\star$ (or critical charge $\tilde{Q}_\star=\sqrt{1-\tilde{a}_\star^2}$) in relation to the extremal rotation $\tilde{a}_{\hbox{\footnotesize ext}}$ (or extremal charge $\tilde{Q}_{\hbox{\footnotesize ext}}$).  When $\tilde{a}_\star<\tilde{a}_{\hbox{\footnotesize ext}}\leq 1$ ($0\leq \tilde{Q}_{\hbox{\footnotesize ext}}< \tilde{Q}_\star$), as is the case in the Kerr limit where $\tilde{a}_{\hbox{\footnotesize ext}}=1$,  the PS family terminates at  $\mathrm{Im} \,\tilde{\omega}=0$ and $\mathrm{Re}\, \tilde{\omega}=m \tilde{\Omega}_H^{\hbox{\footnotesize ext}}$ at extremality.  However, when $\tilde{a}_\star> \tilde{a}_{\hbox{\footnotesize ext}}$ ($\tilde{Q}_\star < \tilde{Q}_{\hbox{\footnotesize ext}}$), as is the case in the RN limit where ${\tilde{Q}_{\hbox{\footnotesize ext}}=1}$, the PS family falls short of the ${(\mathrm{Im} \,\tilde{\omega},\mathrm{Re}\, \tilde{\omega})= (0,m \tilde{\Omega}_H^{\hbox{\footnotesize ext}})}$ surface at extremality. 

Interestingly, the $\star$ transition point turns out to be given (within numerical error) by the point where the separation constant  $\lambda_2(m,\tilde{a}_{\hbox{\footnotesize ext}})$ in \eqref{NH:freq} vanishes:  $\lambda_2(m,\tilde{a}_\star^{\hbox{\tiny NH}})=0$ ($\lambda_2>0$ for $\tilde{a}_{\hbox{\footnotesize ext}}<\tilde{a}_\star^{\hbox{\tiny NH}}$; $\lambda_2<0$ for $\tilde{a}_{\hbox{\footnotesize ext}}>\tilde{a}_\star^{\hbox{\tiny NH}}$).  To get accurate values for $\tilde{a}_\star^{\hbox{\tiny NH}}$ we use the numerical solution for $\lambda_2$. Alternatively, we  get a good approximation by using the WKB result \eqref{NH:lambda2wkb} for $\lambda_2$: 
\be\label{NH:star}
\tilde{a}_\star^{\hbox{\tiny NH}}|_{\hbox{\tiny WKB}}\sim \frac{1}{2}-\frac{5 \sqrt{3} \left(2-\sqrt{2}\right)}{32 \,m}+\frac{5 \left(69-176 \sqrt{2}\right)}{2048 \,m^2}+\mathcal{O}\left(m^{\!-3}\right) 
\ee
In the first case we get $\{\tilde{a}_\star,\tilde{Q}_\star\}^{\hbox{\tiny NH}}\simeq \{0.360, 0.932\}$ while \eqref{NH:star} yields $\{\tilde{a}_\star,\tilde{Q}_\star\}^{\hbox{\tiny NH}}_{\hbox{\tiny WKB}}\sim\{0.311, 0.970\}$ (for $m=2$) \footnote{The eikonal analysis \eqref{PS:eikonal} also predicts the $\star$ transition point. Namely, in the $m\to\infty$ limit it predicts $\tilde{a}_\star^{\hbox{\tiny eikn}}=1/2$ which is, as expected, not a good approximation to $\tilde{a}_\star\simeq 0.360$. However, in the Supplemental Material, we show that the curve $\tilde{a}_\star(m)$  increasingly agrees with the  eikonal PS prediction \eqref{PS:eikonal} when $m$ grows.}.

In summary, our  analysis uncovers a surprising property not observed in the QNM spectra of Schwarzschild, Kerr or RN. Indeed, in the KN QNM spectra we observe a phenomenon know as {\it eigenvalue repulsion} \footnote{It could well be that eigenvalue repulsion is also ultimately responsible for the special features displayed by 1) the $n=5$ (\ie $n=6$ if we start counting them at $n=1$) overtone of the $\ell=m=2$ photon sphere QNM of Kerr (see Fig.~4 of \cite{Onozawa_1997} and \cite{Berti_2009}), and  2) by the QNM  spectra of de Sitter RN black holes \cite{Dias:2020ncd}.}. The latter is common in solid state physics when \eg electrons move in certain Schr\"odinger potentials that introduce energy bands/gaps (see \eg section 7 of~\cite{Kittel:2004}). The eigenvalue repulsion feature is most evident by considering the evolution of the 3 plots in  Fig.~\ref{Fig:spectraFix-a}. In the RN case (left plot), and for small rotation, we have a sharp and unambiguous distinction between the four families of modes represented. In particular,  the $\mathrm{PS}_0$ family dominates the spectra for  $0 \leq \hat{Q}<\hat{Q}_c(\hat{a})$ (with $\hat{Q}_c(0)=\hat{Q}_c^{\hbox{\tiny RN}}$) while the $\mathrm{NH}_0$ family dominates for $\hat{Q}_c(\hat{a})<\hat{Q}\leq 1$. The two modes intersect at $\hat{Q}=\hat{Q}_c(\hat{a})$ with a simple crossover and similar crossovers occur when the $\mathrm{PS}_1$ curve intersects the $\mathrm{NH}_0$ or $\mathrm{NH}_1$ curves. However, at  $a/a_{\hbox{\footnotesize ext}}=0.39$ (middle panel), we find that eigenvalue repulsion occurs between the $\mathrm{PS}_1$ and $\mathrm{NH}_0$ families: the $\mathrm{PS}_1$ curve breaks into two pieces and the same occurs for the $\mathrm{NH}_0$ curve. The left (right) branch of the $\mathrm{PS}_1$  family now connects to the right (left) branch of the $\mathrm{NH}_0$ curve and a {\it frequency gap} appears between the two new curves in the neighbourhood of the two associated kinks. The distinction between the families is no longer sharp. As the rotation increases, new eigenvalue repulsions occur. For example, at $a/a_{\hbox{\footnotesize ext}}=0.96$, the $\mathrm{PS}_0$ curve breaks into two pieces and the same occurs (again) for the $\mathrm{NH}_0$ curve. The left branch of the $\mathrm{PS}_0$ family now merges with the right branch of the $\mathrm{NH}_0$ curve and this new curve is well described by the black curve \eqref{NH:freq} (not shown: the right branch of the  $\mathrm{PS}_0$ curve merges with a $n>1$ NH curve). Below, the left branch of the $\mathrm{NH}_0$ curve now bridges the dark-red triangle $\mathrm{PS}_1$ curve with the blue square $\mathrm{NH}_1$ curve (the $\mathrm{NH}_1$ curve also breaks and merges with another $n>1$ curve but we do not show these further sub-dominant modes).  

\emph{\bf Full QNM spectra.} 
The full spectra of the most dominant KN QNMs  $-$ classified as $Z_2$, $\ell=m=2$, $n=0$ by~\cite{Chandra:1983} (Table V, page 262) in the Schwarzschild limit $-$ is given in Fig.~\ref{Fig:Z2l2m2n0}. The left/right panel gives the imaginary/real part of the frequency. 
The brown curve has $\mathrm{Im}\,\tilde{\omega}=0$, $\mathrm{Re}\,\tilde{\omega}=m\tilde{\Omega}_H^{\hbox{\footnotesize ext}}$. 
To scan the 2-dimensional parameter space we used a grid with $100\times 100$ points in $[0,1]\times[0,1]$ for $\{\hat{Q}, a/a_{\hbox{\footnotesize ext}}\}$ with $\hat{a}_{\hbox{\footnotesize ext}} = \sqrt{1-\hat{Q}^2}$.

The KN modes with slowest decay rate always terminate at extremality along the extremal brown curve, with the frequencies off-extremality well approximated by \eqref{NH:freq} as illustrated in Fig.~\ref{Fig:spectraFix-a}.
The red surface family, continuously connected to the Schwarzschild mode (dark-red point ~\cite{Chandra:1983,Leaver:1985ax}), is the $\mathrm{PS}_0$ QNM family as we unambiguously identify it in the RN limit. It dominates the spectra for most of the parameter space. However, for large $\tilde{Q}$ it is instead  the $\mathrm{NH}_0$ QNM family (green surface) that has the  lowest $|\mathrm{Im}\,\tilde{\omega}|$. In between these orange/green regions there is a yellowish zone. This is where either simple crossovers  (that trade mode dominance) or eigenvalue repulsions between the $\mathrm{PS}_0$ and $\mathrm{NH}_0$ modes occurs. These were already analysed in the discussion of Fig.~\ref{Fig:spectraFix-a}.
The derived QNM spectra can be used to model beyond Standard Model physics in binary mergers and GW emission in realistic astrophysical environments, bearing increasing importance with future enhancements in sensitivity of current and planned GW observatories. 
In a companion paper \cite{AstroConstraints}, we apply the results obtained in this work to the latest observations from the GW detector network.


\smallskip


\noindent{\bf Acknowledgments.}

The authors would like to thank Nathan Johnson-McDaniel for helpful discussions.
We acknowledge the use of the cluster `Baltasar-Sete-S\'ois', and associated support services at CENTRA/IST, in the completion of this work. The authors further acknowledge the use of the IRIDIS High Performance Computing Facility, and associated support services at the University of Southampton, in the completion of this work.
O.~C.~D. acknowledges financial support from the STFC ``Particle Physics Grants Panel (PPGP) 2016" Grant No.~ST/P000711/1 and the STFC ``Particle Physics Grants Panel (PPGP) 2018" Grant No.~ST/T000775/1. M.~G is supported by a Royal Society University Research Fellowship. J.~E.~S. has been partially supported by STFC consolidated grants ST/P000681/1, ST/T000694/1. The research leading to these results has received funding from the European Research Council under the European Community's Seventh Framework Programme (FP7/2007-2013) / ERC grant agreement no. [247252]. 

\newpage
\begin{appendix}
\onecolumngrid
\section{\Large Supplemental Material}\label{sec:Appendix}

\subsection{Coupled pair of PDEs for the KN perturbations}\label{sec:Appendix1}

The uniqueness theorems \cite{Robinson:2004zz,Chrusciel:2012jk} state that the Kerr-Newman (KN) black hole (BH) is the unique, most general family of stationary asymptotically flat BHs, of Einstein-Maxwell theory. It is characterised by 3 parameters: mass $M$, angular momentum $J\equiv M a$ and charge $Q$. The Kerr, Reissner-Nordstr\"om (RN) and Schwarzschild (Schw) BHs constitute limiting cases: ${Q=0}$, $a=0$ and $Q=a=0$, respectively.
The gravitational and Maxwell fields of the KN BH in  Boyer-Lindquist coordinates are given by \cite{Newman:1965my,Adamo:2014baa}
\begin{eqnarray}\label{KNsoln}
ds^2&=&-\frac{\Delta}{\Sigma} \left(\dd t-a \sin^2\theta \dd \phi  \right)^2+\frac{\Sigma }{\Delta }\,\dd r^2 + \Sigma \,\dd \theta^2 
+ \frac{\sin ^2\theta}{\Sigma }\left[\left(r^2+a^2\right)\dd \phi -a \dd t \right]^2, \nonumber\\
A&=& \frac{Q \,r}{\Sigma}\left(\dd t-a \sin^2\theta \dd \phi \right),
\end{eqnarray}
with $\Delta = r^2 -2Mr+a^2+Q^2$ and $\Sigma=r^2+a^2 \cos^2\theta$. 

Linear gravito-electromagnetic perturbations about the KN background are more easily addressed in the  Newman-Penrose (NP) formalism \cite{Newman:1961qr}. 
In the context of this formalism there is a well-known set of NP scalars built of contractions of the NP tetrad with the Weyl tensor (\eg $\Psi_2$, $\Psi_3$ and $\Psi_4$) or with the Maxwell field strength (\eg $\Phi_1$ and $\Phi_2$) \cite{Chandra:1983,Stephani:2003tm} . Out of these, one can construct two  
{\it gauge invariant} perturbed quantities, \ie quantities that are invariant under both linear diffeomorphisms and tetrad rotations, namely  \cite{Dias:2015wqa}:
\begin{eqnarray}\label{gauging}
&&\psi_{-2}= \left(\bar{r}^*\right)^4 \Psi_4^{(1)},  \nonumber\\
&& \psi_{-1}=\frac{\left(\bar{r}^*\right)^3}{2\sqrt{2}\Phi_1^{(0)}} \left(2\Phi_1^{(0)}\Psi_3^{(1)} -3 \Psi_2^{(0)}\Phi_2^{(1)}\right),
\end{eqnarray}
with $\bar{r} = r+ia\cos \theta$. Here, NP scalars with superscript $^{(0)}$ refer to scalars in the KN background and the superscript $^{(1)}$ to first order perturbations of the scalar.  
These  NP scalars \eqref{gauging} are the ones relevant for the study of perturbations that are outgoing at future null infinity and regular at the future horizon \footnote{There is a set of two coupled PDEs---related to (4) by a Geroch-Held-Penrose \cite{Geroch:1973am} transformation---for the quantities $\psi_{2}$ and $\psi_{1}$ that are the positive spin counterparts of (4); however these would be relevant if we were interested in perturbations that were outgoing at past null infinity.}.  
Ref. \cite{Dias:2015wqa} derived a set of two coupled partial differential equations (PDEs) for $\psi_{-2}$ and $\psi_{-1}$ that describe the most general perturbations (except for trivial modes that shift the parameters of the solution) of a KN BH, namely:
\begin{eqnarray}\label{ChandraEqsAppendix}
&& \left(\mathcal{F}_{-2}+ Q^2 \mathcal{G}_{-2}\right) \psi_{-2}  + Q^2 \mathcal{H}_{-2}  \psi_{-1} =0 \,, \nonumber \\
&& \left(\mathcal{F}_{-1} +Q^2 \mathcal{G}_{-1}\right)\psi_{-1} + Q^2  \mathcal{H}_{-1} \psi_{-2}=0  \,, 
\end{eqnarray}
where the second order differential operators $\{\mathcal{F},\mathcal{G},\mathcal{H}\}$ are given by  \cite{Dias:2015wqa}
\begin{eqnarray}\label{def:opsFGH}
\mathcal{F}_{-2}&=&\Delta\DD_{-1}^\dagger\DD_0 +\LL_{-1}\LL_2^\dagger -6i \omega\bar{r} \,,
\nonumber \\
\mathcal{G}_{-2}&=&\Delta \DD_{-1}^\dagger \alpha_-\bar{r}^*\DD_0  -3\Delta \DD_{-1}^\dagger \alpha_- 
- \LL_{-1}\alpha_+ \bar{r}^* \LL_2 ^\dagger  +3 \LL_{-1} \alpha_+ i a \sin \theta \,, 
\nonumber \\
\mathcal{H}_{-2}&=&-\Delta \DD_{-1}^\dagger \alpha_- \bar{r}^* \LL_{-1}  -3 \Delta \DD_{-1}^\dagger \alpha_- i a \sin \theta 
-\LL_{-1} \alpha_+ \bar{r}^* \Delta \DD_{-1}^\dagger  -3\LL_{-1} \alpha_+ \Delta  \,,
\nonumber \\
\mathcal{F}_{-1}&=&\Delta\DD_1\DD_{-1}^\dagger +\LL_2^\dagger\LL_{-1}-6i \omega\bar{r} \,,
\\
\mathcal{G}_{-1}&=& - \DD_0 \alpha_+ \bar{r}^* \Delta \DD_{-1}^\dagger  -3 \DD_0 \alpha_+ \Delta 
 +\LL_2^\dagger \alpha_- \bar{r}^* \LL_{-1}  +3 \LL_2^\dagger \alpha_- i a\sin\theta \,,
\nonumber \\
\mathcal{H}_{-1}&=& -\DD_0 \alpha_+ \bar{r}^* \LL_2^\dagger  +3 \DD_0 \alpha_+ i a \sin \theta  
 -\LL_2^\dagger \alpha_- \bar{r}^* \DD_0  +3 \LL_2^\dagger \alpha _- \,,
\nonumber
\end{eqnarray}
with 
$\alpha_\pm \equiv \left[3(\bar{r}^2M-\bar{r} Q^2)\pm Q^2\bar{r}^*\right]^{-1}$,  and we introduced the radial and angular Chandrasekhar operators  \cite{Chandra:1983},
\begin{eqnarray}\label{def:DL}
&& \DD_j = \partial_r+\frac{i K_r}{\Delta}+2j\frac{(r-M)}{\Delta}, \quad K_r=am-(r^2+a^2)\omega; \nonumber \\
 && \LL_j = \partial_\theta+K_{\theta}+j\cot\theta, \quad K_{\theta}=\frac{m}{\sin\theta}-a\omega\sin\theta. 
\end{eqnarray}
The complex conjugate of these operators, namely $\DD_j^\dagger $ and $\LL_j^\dagger$, can be obtained from $\DD_j$ and $\LL_j$ via the replacement $K_r \to - K_r$ and $K_{\theta} \to - K_{\theta}$, respectively.

Note that fixing a gauge in which $\Phi_{0}^{(1)}=\Phi_{1}^{(1)}=0$, \eqref{ChandraEqsAppendix} reduces to the Chandrasekhar coupled PDE system \cite{Chandra:1983} (see also the derivation in \cite{Mark:2014aja}).  Finally, note that in the limit $Q\to 0$ \eqref{ChandraEqsAppendix}  decouple yielding the familiar Teukolsky equation for Kerr  \cite{Teukolsky:1972my}.

Since $\partial_t,\partial_\phi$ are Killing vector fields of KN, we can Fourier decompose the perturbations $\{\psi_{-2},\psi_{-1}\}$ as $e^{-i \omega t} e^{i m \phi}$. This introduces the frequency $\omega$ and azimuthal quantum number $m$ of the perturbation. The $t -\phi$ symmetry of the KN BH allows to consider only modes with  Re$(\omega)\geq 0$, as long as we study both signs of $m$. 
Then, to solve the coupled PDEs \eqref{gauging}, we need to impose physical boundary conditions (BCs). 
At spatial infinity, a Frobenius analysis of \eqref{ChandraEqsAppendix} that allows only outgoing waves yields the decay:
\begin{equation}\label{BC:inf}
\psi_{s}{\bigl |}_\infty \!\simeq\!e^{i \omega r} r^{-(2s+1)+i \omega \,\frac{r_+^2+a^2+Q^2}{r_+}} \!\! \left( \!\! \alpha_{s}(\theta)+\frac{\beta_{s}(\theta)}{r}+\cdots\!\!\right), \nonumber
\end{equation}
where $s=-2,-1$, and $\beta_{s}(\theta)$ is a function of $\alpha_{s}(\theta)$ and its derivative fixed by expanding \eqref{ChandraEqsAppendix} at spatial infinity.

At the horizon, a Frobenius analysis whereby we require only regular modes in ingoing Eddington-Finkelstein coordinates, yields the expansion 
\begin{equation}\label{BC:H}
\psi_{s}{\bigl |}_H \!\simeq\! \left(r-r_+\right)^{-s-\frac{i (\omega -m\Omega_H)}{4 \pi T_H}}[ a_{s}(\theta)+ b_{s}(\theta)(r-r_+) +\cdots ], \nonumber
\end{equation}
where $b_{s}(\theta)$ is a function of $a_s(\theta)$ and its derivative.

At the North (South) pole $x\equiv \cos\theta =1\,(-1)$, regularity dictates that the fields must behave as 
($\varepsilon = 1$ for $|m|\geq 2$, while $\varepsilon = -1$ for $|m|=0,1$ modes)
\begin{equation}\label{BC:N}
\psi_{s}{\bigl |}_{\hbox{\tiny N,(S)}} \hspace{-1pt} \simeq  (1\mp x)^{\varepsilon^{\frac{1\pm 1}{2}} \frac{s+|m|}{2}} \hspace{-1pt} \left[ A^{\pm}_{s}(r)+B^{\pm}_{s}(r)(1\mp x)+\cdots \right],  \nonumber
\end{equation}
where $B^{+}_{s}(r)$($B^{-}_{s}(r)$) is a function of $A^{+}_{s}(r)$($A^{-}_{s}(r)$) and its derivatives along $r$, whose exact form is fixed by expanding \eqref{ChandraEqsAppendix} around the North (South) pole.

\subsection{WKB coefficients for the separation constant $\lambda_2$}\label{sec:NHappendix}

At extremality, the modes with slowest decay rate (independently of belonging to the NH or PS families) always approach $\mathrm{Im}\,\tilde{\omega}=0$ and $\mathrm{Re}\,\tilde{\omega}=m \tilde{\Omega}_H^{\hbox{\footnotesize{ext}}}$ and \eqref{NH:freq} of the main text provides an excellent approximation to their frequency in an expansion off-extremality (as analysed in the discussion of Fig.~\ref{Fig:spectraFix-a} of the main text). The derivation of the analytical approximation \eqref{NH:freq} of the main text is quite long and thus we will present it in the companion manuscript \cite{ExtendedQNMsKN}.

In \eqref{NH:freq} of the main text, the separation constant $\lambda_2$ has a WKB expansion for large $m$, as given in Eq. \eqref{NH:lambda2wkb} of the main text. The associated WKB coefficients 
are:
\begin{subequations}\label{NH:WKBansatzCoef}
\begin{align}
&\lambda_{2,0}=4 \left(1-4 \hat{a} ^2\right),\qquad 
\lambda_{2,1}= -4 \left(1+\hat{a}^2\right) \left(2 \sqrt{1-\hat{a} ^2}-\sqrt{1+2 \hat{a} ^2}\right),
   \\
&\lambda_{2,2}= \frac{3 \sqrt{1-\hat{a} ^2} \left(1+\hat{a}^2\right)^2 \left(3-726 \hat{a} ^{10}-253 \hat{a} ^8+128 \hat{a} ^6-74 \hat{a} ^4-50 \hat{a} ^2\right)}{\left(1+2\hat{a}^2\right) \left[\left(66 \hat{a} ^6-5 \hat{a} ^4-12 \hat{a} ^2+5\right) \sqrt{1-\hat{a} ^2}+4 \left(1-\hat{a} ^4\right) \sqrt{2 \hat{a} ^2+1}\right]},  
 \tag{\stepcounter{equation}\theequation}   \\
&\lambda_{2,3}=
\bigg[ 
4 \left(1+2\hat{a}^2\right)^{7/2}\bigg(
578577650112 \hat{a} ^{40}-338129795520 \hat{a} ^{38}-1042453021104 \hat{a} ^{36}+1170932108544 \hat{a} ^{34}
\nonumber \\
&\hspace{0.8cm} +243872180244 \hat{a} ^{32}-1092788709804 \hat{a} ^{30}+457571937931 \hat{a} ^{28}+286639850738 \hat{a} ^{26}-371225227587 \hat{a} ^{24}
\nonumber \\
&\hspace{0.8cm} 
+75821376048 \hat{a} ^{22}+83823143199 \hat{a} ^{20}-64522516578 \hat{a} ^{18}+5397537793 \hat{a} ^{16}+11870759300 \hat{a} ^{14}-5939331087 \hat{a} ^{12}
\nonumber \\
&\hspace{0.8cm} 
+15670254 \hat{a} ^{10}+798959271 \hat{a} ^8-269248008 \hat{a} ^6-8868395 \hat{a} ^4+20327618 \hat{a} ^2-4782969
\bigg)
\nonumber \\
&\hspace{0.8cm} 
+4 \sqrt{1-\hat{a} ^2} \left(1+2\hat{a}^2\right)^3\bigg(
661231600128 \hat{a} ^{40}-788969522880 \hat{a} ^{38}-475886378880 \hat{a} ^{36}+1029138506352 \hat{a} ^{34}
\nonumber \\
&\hspace{0.8cm} 
-630648141552 \hat{a} ^{32}-452699156052 \hat{a} ^{30}+658166339168 \hat{a} ^{28}-186975958943 \hat{a} ^{26}-249892000005 \hat{a} ^{24}
\nonumber \\
&\hspace{0.8cm} 
+178743692406 \hat{a} ^{22}-3249242106 \hat{a} ^{20}-56479482309 \hat{a} ^{18}+20902690721 \hat{a} ^{16}+3663601312 \hat{a} ^{14}-5845481340 \hat{a} ^{12}
\nonumber \\
&\hspace{0.8cm} 
+1100552199 \hat{a} ^{10}+410656173 \hat{a} ^8-279409506 \hat{a} ^6+19829366 \hat{a} ^4+13153165 \hat{a} ^2-4782969 
\bigg)\bigg]^{-1}
\nonumber \\
&\hspace{0.8cm} 
\bigg[ 
3 \hat{a} ^2 \sqrt{1-\hat{a} ^2} \left(1+\hat{a}^2\right)^3 \sqrt{2 \hat{a} ^2+1} \bigg(
90588729217536 \hat{a} ^{46}+93586813404480 \hat{a} ^{44}-64234642488192 \hat{a} ^{42}
\nonumber \\
&\hspace{0.8cm} 
-54181551934224 \hat{a} ^{40}+14733709326864 \hat{a} ^{38}-34708141099764 \hat{a} ^{36}-8979094220672 \hat{a} ^{34}+34432474064505 \hat{a} ^{32}
\nonumber \\
&\hspace{0.8cm}
-10922161747605 \hat{a} ^{30}-23041644949212 \hat{a} ^{28}+5136927583340 \hat{a} ^{26}+4733507876355 \hat{a} ^{24}-3578226571619 \hat{a} ^{22}
\nonumber \\
&\hspace{0.8cm} 
-898929274206 \hat{a} ^{20}+753565243446 \hat{a} ^{18}-135077374365 \hat{a} ^{16}-174223122235 \hat{a} ^{14}+33089919120 \hat{a} ^{12}
\nonumber \\
&\hspace{0.8cm} 
+8380363168 \hat{a} ^{10}-9890782275 \hat{a} ^8-803782461 \hat{a} ^6+541670718 \hat{a} ^4-148272034 \hat{a} ^2-57395628  \bigg)
\nonumber \\
&\hspace{0.8cm} 
+3 \hat{a} ^2 \left(1+\hat{a}^2\right)^3 \bigg( 
158530276130688 \hat{a} ^{48}+192260601732672 \hat{a} ^{46}-226279077675552 \hat{a} ^{44}
\nonumber \\
&\hspace{0.8cm} 
-257580189150768 \hat{a} ^{42}+238634465705064 \hat{a} ^{40}+187478664334236 \hat{a} ^{38}-167948153974214 \hat{a} ^{36}
\nonumber \\
&\hspace{0.8cm} 
-79050787933609 \hat{a} ^{34}+69165996968940 \hat{a} ^{32}+1562277529575 \hat{a} ^{30}-26149776558142 \hat{a} ^{28}
\nonumber \\
&\hspace{0.8cm} 
+6310859786413 \hat{a} ^{26}+3820171951948 \hat{a} ^{24}-4424582883901 \hat{a} ^{22}-417658252182 \hat{a} ^{20}+868831525263 \hat{a} ^{18}
\nonumber \\
&\hspace{0.8cm} 
-249677209480 \hat{a} ^{16}-170706582299 \hat{a} ^{14}+47404470046 \hat{a} ^{12}+4708012127 \hat{a} ^{10}-10932078636 \hat{a} ^8-398469675 \hat{a} ^6
\nonumber \\
&\hspace{0.8cm} 
+532105820 \hat{a} ^4-176969858 \hat{a} ^2-57395628
\bigg)\bigg].
\end{align}
\end{subequations}
The derivation of \eqref{NH:lambda2wkb} of the main text and of \eqref{NH:WKBansatzCoef} is again  long and will be given it in the companion manuscript \cite{ExtendedQNMsKN}. There, we also show that this WKB expansion provides an excellent approximation already for $m=10$ and a good approximation even for $m=2$. 

\end{appendix}

\bibliography{refsKN}{}
\bibliographystyle{apsrev}

\end{document}